\documentclass[aps,prl,preprint,twocolumn,superscriptaddress,10pt]{revtex4-1}
\usepackage{graphicx}

\begin{document}

\title{Distillation of the two-mode squeezed state}

\author{Yury Kurochkin}
\affiliation{Russian Quantum Centre, 100 Novaya St., Skolkovo,
Moscow 143025, Moscow, Russia}
\author{Adarsh S. Prasad}
\affiliation{Institute for Quantum Science and Technology, University of Calgary, Alberta T2N1N4, Canada}

\author{A. I. Lvovsky}
\affiliation{Russian Quantum Centre, 100 Novaya St., Skolkovo,
Moscow 143025, Moscow, Russia}
\affiliation{Institute for Quantum Science and Technology, University of Calgary, Alberta T2N1N4, Canada}
\email[]{LVOV@ucalgary.ca}

\newcommand{\bra}[1]{\left\langle #1\right|}
\newcommand{\ket}[1]{\left| #1\right\rangle}
\newcommand{\braket}[2]{\left\langle
#1\vphantom{#2}\right|\left.#2\vphantom{#1}\right\rangle}
\newcommand{\ketbra}[2]{\left| #1\right\rangle\!\left\langle#2\right|}
\newcommand{\avg}[1]{\left\langle #1\right\rangle}
\newcommand{\be}[0]{\begin{equation}}
\newcommand{\ee}[0]{\end{equation}}
\newcommand{\intinf}[0]{\int_{-\infty}^{+\infty}}
\newcommand{\ada}[0]{\hat a^\dagger\hat a}
\newcommand{\ad}[0]{\hat a^\dagger}
\newcommand{\ah}[0]{\hat a}
\newcommand{\hv}[1]{\hat{\vec{#1}}}
\newcommand{\tr}[0]{{\rm Tr}}
\newcommand{\re}[0]{{\rm Re}\,}
\newcommand{\im}[0]{{\rm Im}\,}
\newcommand{\lra}\leftrightarrow
\newcommand{\eqref}[1]{(\ref{#1})}
\newcommand{\eeqref}[1]{Eq.~(\ref{#1})}
\newcommand{\braketop}[3]{\left\langle
#1\vphantom{#2#3}\right|\left.#2\vphantom{#1#3}\right|\left.#3\vphantom{#1#2}\right\rangle}
\newcommand{\expec}[1]{\left\langle #1\right\rangle}
\newcommand{\pr}[0]{{\rm pr}}
\newcommand{\zero}[0]{\ket{{\rm zero}}}
\newcommand{\adj}[1]{{\rm{Adjoint}}\left({#1}\right)}
\newcommand{\mat}[2]{\left(\begin{array}{#1} #2 \end{array}\right)}
\newcommand{\de}[0] {{\rm d}}
\newcommand{\os}[1] {\overset{\eqref{#1}}{=}}
\newcommand{\nna}[0] {\nonumber \\}
\newcommand{\nnb
}[0] {\\ \nonumber }
\newcommand{\iea}[0]{{\it et al.~}}
\newcommand{\ieac}[0]{{\it et al., }}
\newcommand{\sq}[0]{\ket{{\rm sq}_R}}
\newcommand{\tmsv}[0]{\ket{{\rm TMSV}_R}}
\begin{abstract}
We experimentally demonstrate entanglement distillation of the two mode squeezed state. Applying the photon annihilation operator to both modes, we raise the fraction of the two-photon component in the state, resulting in the increase of both squeezing and entanglement by about 50\%.
\end{abstract}

\date{\today}



\maketitle




Entanglement is paramount in quantum technology. However, entangled states are difficult to prepare and vulnerable to decoherence and losses. This issue is particularly significant in quantum optical communications where entanglement needs to be distributed over long distances.

It can be addressed using entanglement distillation (ED) \cite{bennett96}, a procedure in which the parties use classical communications and local operations to obtain, from a set of entangled states, a smaller set of states with a higher level of entanglement. ED has been successfully demonstrated in the discrete-variable domain \cite{kwiat, pan}. But for continuous variable (CV) quantum information processing, ED is complicated because of a no-go theorem \cite{PRL89_137903_2002, Eisert_no_go, Cirac_no_go} that prohibits distillation of a Gaussian entangled state by any Gaussian operations. Gaussian operations include phase-space displacement, squeezing, application of beam splitters, homodyne detection --- i.e. such operations that preserve the Gaussian character of a state's Wigner function and are typical for CV processing. The primary continuous-variable entangled resource, the two-mode squeezed vacuum (TMSV), is Gaussian \cite{Braunstein_review}, so one must leave the boundaries of the the CV domain in order to distill it. This is the purpose of the present work.

TMSV is of special value for quantum science and technology \cite{Braunstein_review}. Thanks to nonclassical correlation of quadrature observables of the two modes, this state constitutes a physically plausible approximation of the original Einstein, Podolsky and Rosen state \cite{Einstein35} that triggered the quantum nonlocality debate. In addition to fundamental interest, TMSV is the basis of complete quantum teleportation \cite{furusawa1998} and CV quantum repeaters \cite{Campbell13} as well as certain quantum metrology \cite{Anisimov10,Carranza12} and quantum key distribution \cite{Madsen12,GG02} applications. Hence it is important to develop and test a reliable procedure for the distillation of that state.


A non-Gaussian operation that has been extensively discussed in the context of CV ED is the photon annihilation operator $\hat a$. It is realized by reflecting the target state $\ket\psi$ off a low transmissivity beamsplitter with the single photon detector placed in its transmitted channel. If the detector registers a photon, the state reflected from the beam splitter is the photon subtracted state $\hat a\ket\psi$. This operation has been first implemented by Wenger \iea \cite{Wenger04} and has since been used in a number of CV quantum engineering experiments \cite{lvovsky2009,Kumar13}.

Entanglement distillation with photon annihilation was first proposed by Opartn\'{y} \textit{et al}. \cite{Opartny} and further theoretically investigated for photon-number resolving \cite{Cochrane02} and threshold detectors \cite{Olivares03}. A comprehensive theoretical analysis in Ref.~\cite{oxford2013} considered different types of detectors and realistic noisy measurements. The utility of TMSV distilled by dual-mode photon annihilation for teleportation of highly nonclassical states has been discussed in Ref.~\cite{Agarwal13}.

CV entanglement increase by photon annihilation has been demonstrated experimentally, but in none of the existing experiments did the resulting state retain the two-mode squeezing property. Ourjoumtsev  \textit{et al}. applied non-local photon subtraction to TMSV, resulting in a state that approximates the delocalized single photon \cite{ourjoumtsev2007,ourjoumtsev2009}. CV ED employing two-mode photon annihilation was demonstrated for Gaussian input states created by equal splitting of a single mode squeezed state in \cite{takahashi2010}. In that work, entanglement increase has been observed along with enhanced non-classical quadrature correlations for specific values of the phase difference between the two modes. In another set of experiments, artificial non-Gaussian disturbance has been applied to a TMSV, resulting in a loss of its Gaussian character. Subsequently, the entanglement was distilled by means of a Gaussian process \cite{Hage08,Dong08}, but not beyond the entanglement level of the original TMSV.




We now proceed to explaining the idea of our work in an idealized, loss-free setting. TMSV is obtained by applying the two-mode squeezing operator $S(\zeta) = e^{\zeta (\hat{a_1} \hat{a_2} - \hat{a}^\dag_1 \hat{a}^{\dag}_2)}$ (where $\zeta$ is the real squeezing parameter) to modes 1 and 2 initially in the vacuum state. It has the following representation in the photon number basis:
\begin{equation}\label{Psiin}
\ket{\Psi}= S(\zeta)|0,0\rangle=\sqrt{1-\lambda^2}\sum^{\infty}_{n=0} \lambda^n |n,n\rangle,
\label{eq:tmss}
\end{equation}
for $\lambda = \tanh\zeta$. If we apply  annihilation operators to both modes of $\ket{\Psi}$, we find
\begin{equation}\label{Psiout}
\hat{a}_1 \hat{a}_2 \ket\Psi =\sqrt{1-\lambda^2} \sum^{\infty}_{n=1} n\lambda^{n} |n-1,n-1\rangle
\end{equation}
In the limit of small squeezing, both states \eqref{Psiin} and \eqref{Psiout} can be approximated to its first order. After renormalization, we find
\begin{eqnarray}
\ket\Psi|&\approx&|0,0\rangle+\lambda |1,1\rangle \\
\hat{a}_1 \hat{a}_2 \ket\Psi&\approx&|0,0\rangle+2\lambda |1,1\rangle
\label{eq:psiin}
\end{eqnarray}
A higher contribution of the double-photon term causes the entanglement increase \cite{oxford2013}. In addition, a higher $\lambda$ manifests itself through increase of two-mode squeezing, i.e. enhanced nonclassical correlation of quadrature observables of the two modes.

In the ideal case, entanglement distillation is also expected if photon annihilation is applied to only one of the modes of the initial TMSV:
\begin{equation}
\hat{a}_1 \ket\Psi\approx |0,1\rangle+\sqrt{2} \lambda |1,2\rangle
\label{eq:psione}
\end{equation}
However, this state no longer has the form of TMSV and is not expected to feature nonclassical quadrature correlation. Furthermore, as we see below, in our experiment, state $\hat{a}_1 \ket\Psi$ does not exhibit increased entanglement because of the losses.

A signature feature of TMSV is the correlation of quadrature measurement statistics. In the two modes, the position and momentum observables are, respectively, correlated and anticorrelated beyond the level allowed by the uncertainty principle for separable states:
\begin{eqnarray}
\langle (X_1 \mp X_2)^2 \rangle &=&e^{\mp 2\zeta}; \nna
\langle (P_1 \mp P_2)^2 \rangle &=&e^{\pm 2\zeta}. \nonumber
\end{eqnarray}
For quadrature observables $X(\theta)=X \cos{\theta} + P \sin{\theta}$ associated with arbitrary phases in the two modes, the correlated variance takes the form
\begin{eqnarray}
\langle [X_1(\theta_1) \mp X_2(\theta_2)]^2 \rangle=&
\label{eq:phase_dependence}
\\ &\hspace{-3cm}=  \{(1-\eta) + \eta[\cosh(2\zeta)\pm \cos(\theta_1+\theta_2)\sinh(2\zeta)] \} \nonumber
\end{eqnarray}
where we have taken into account the optical loss $1-\eta$ in both modes. Remarkably, the correlation depends only on the sum but not the difference of the  phases in the two modes. This feature can be understood by reviewing the photon number decomposition \eqref{Psiin} of TMSV. A phase shift by angle $\theta$ in a single mode corresponds to operator $e^{i\theta\hat n}$, where $\hat n$ is the photon number operator. Because all terms of \eeqref{Psiin} contain equal photon numbers in both modes, a change in $\theta_1-\theta_2$ for constant $\theta_1+\theta_2$ implies an equal and opposite quantum phase shift of the two modes, and will leave the state unchanged. This characteristic feature of TMSV is preserved in our ED protocol, in contrast to previous photon subtraction CV ED experiments \cite{ourjoumtsev2007,ourjoumtsev2009,takahashi2010}.

\begin{figure}[h]
\includegraphics[width=0.4\textwidth]{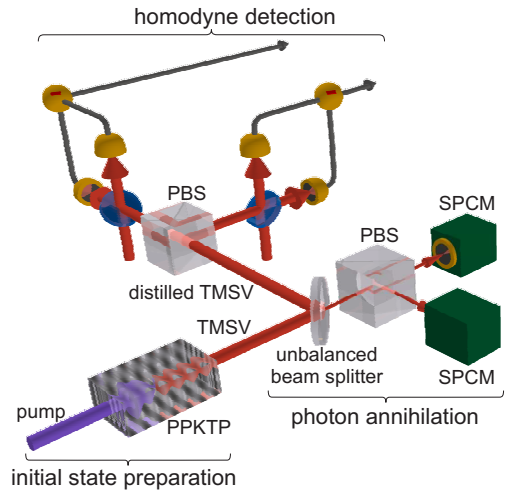}
\caption{\label{fig:exp_setup} Experimental setup. Type II parametric down conversion in a PPKTP crystal generates the two-mode squeezed vacuum state in two polarization modes. Each mode is subjected to the annihilation operator realized by an unbalanced beamsplitter and a single photon detector. The prepared distilled two-mode squeezed state has higher squeezing and entanglement, as verified by homodyne detection.}
\end{figure}

In our experiment (Fig.~\ref{fig:exp_setup}), we generate a TMSV using nondegenerate parametric down-conversion and perform multiple quadrature measurements in both modes of the original and distilled states. The acquired data allow us to verify entanglement distillation in two ways. First, we evaluate the phase-dependent variance of the sum and difference of the quadratures acquired in the two modes and verify that the minimal value of that variance decreases after distillation, corresponding to a higher squeezing. Second, we use the complete set of quadrature data for full characterization of the original and distilled states by means of homodyne tomography \cite{lvovsky2009}. We then verify the entanglement increase by evaluating, for both states, the log-negativity value $E_N = \log_2(1+2N)$, where negativity $N$ is the entanglement monotone  \cite{Plenio05} equal to the sum of absolute values of negative eigenvalues of the state's partially transposed density matrix. For Gaussian states, the log-negativity is a proper measure of entanglement \cite{Plenio03}.

\begin{figure*}[ht]
\includegraphics[width=\textwidth]{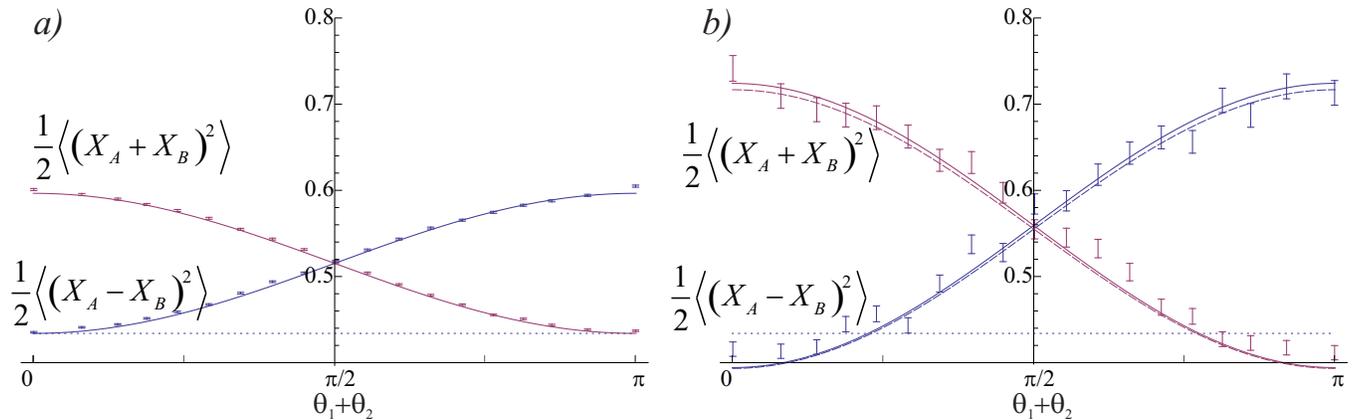}
\caption{\label{fig:XmXpN} Variances of the sum and difference of the quadratures measured in the two modes: a) unconditionally, b) conditioned on photon annihilation events in both channels. The noise level corresponding to the double vacuum state is 1. The minimum variance in (a) is indicated by a dashed line. The solid line in (a) is the fit based on a lossy TMSV model with the squeezing parameter of $\zeta=0.19$; the solid line in (b) is a theoretical prediction based on the fit in (a) and photon annihilation applied in both modes. The dashed line in (b) corresponds to a TMSV model with increased squeezing parameter $\zeta=0.358$ that has experienced the same loss as the state in (a). The dotted line in both panes corresponds to the highest level of squeezing in the unconditionally measured state, according to the fit.}
\end{figure*}

The experimental setup is presented on Fig.~\ref{fig:exp_setup}.
The two-mode squeezed state is prepared by a type II spontaneous parametric downconversion in a PPKTP crystal in a spatially and spectrally degenerate but polarization-nondegenerate configuration. The PPKTP crystal is pumped by 395-nm wavelength pulses generated by doubling 790-nm pulses from a master Ti:sapphire mode locked laser, with a repetition rate of 76 MHz and pulse width 1.6 ps.  The down-conversion is followed by a polarization independent beamsplitter with 11\% transmissivity. The transmitted signal is subjected to narrowband spectral filtering, after which it is separated according to polarization and each mode is subjected to single photon detection. PerkinElmer SPCM-AQR-14-FC detectors, coupled through single-mode fibers, are used \cite{Huisman09}. In spite of non-negligible two-mode squeezing and an over 50\% efficiency of the detectors, the coincidence photon count rate is only $\sim 100$ Hz because of the losses associated with the spatial and spectral filtering \footnote{$\gamma^2\approx0.04$ photon pairs per pulse are produced at a 76 MHz pulse repetition rate. Each photon passes the photon subtraction beamsplitter with 11\% probability. The spatial and spectral filters account for additional $\sim 90\%$ losses for each photon, and the single photon detector efficiency equals $~60\%$. This results in coincidence clicks on a scale of 100 Hz}. The light reflected from the beamsplitter is separated into the two TMSV modes by a polarizing beam splitter and each mode is subjected to homodyne
measurements \cite{Kumar12}. The local oscillators for the balanced homodyne detectors are derived from the master laser.

Photon annihilation events correspond to ``clicks" of photon detectors in the relevant mode. Events without detector clicks correspond to the initial TMSV $\ket\Psi$
and are used to measure the local oscillator phases as well as quantify initial squeezing and entanglement. Simultaneous clicks in both detectors herald the distilled TMSV state, for which we observe the squeezing and entanglement increase.

A detector click in only one of the modes (e.g. mode 1) leads to the one-photon subtraction state (\ref{eq:psione}) which approximates, in the limit of low squeezing, a tensor product of the vacuum and single-photon states. In the presence of losses, the state of mode 2 becomes a mixture of the single-photon and vacuum states. Reconstructing the state of that mode permits precise evaluation of the loss present in the experiment as  $1-\eta=0.58 \pm 0.01$ \cite{Lvovsky_single_photon}. This figure includes the 11\% loss from the beamsplitter used for photon subtraction.

\begin{figure}[hb]
\includegraphics[width=0.3\textwidth]{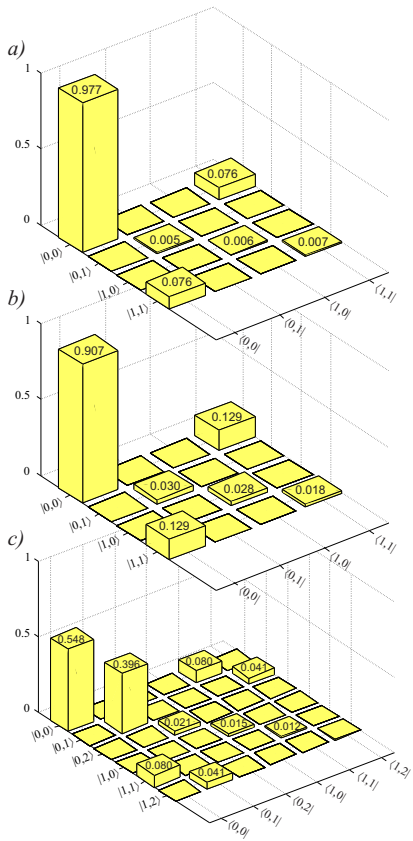}
\caption{\label{fig:rho}Low photon number components of the density matrices (absolute values) reconstructed from the quadrature data sets measured: a) unconditionally, b) conditioned on photon annihilation events in both channels, c) conditioned on photon a annihilation event in channel 1. Increase in the off-diagonal elements associated with terms $\ketbra{0,0}{1,1}$ and  $\ketbra{1,1}{0,0}$ in (b) compared to (a) is evidence of entanglement distillation. All matrix elements above 0.005 are marked. No compensation for the loss is implemented.}
\end{figure}

We allow the local oscillator phase in one of the homodyne detectors to fluctuate freely with air movements while that in the other channel is varied with a period of about 1 s by means of a piezoelectric transducer. With each heralding event, we acquire a single pair of quadratures associated with that event, plus a series of 9500  quadrature pairs associated with subsequent (not heralded) laser pulses. We can safely assume the local oscillator phases to be constant during that acquisition. Because the output of the parametric down-conversion is a two-mode squeezed state, the correlated quadrature variance $\langle (X_1 \mp X_2)^2 \rangle$ in each series provides us with the information on the sum of the local oscillator phases $(\theta_1 + \theta_2)$ in accordance with Eq.~(\ref{eq:phase_dependence}).

A plot of measured variance of $\langle (X_1 \mp X_2)^2 \rangle$ versus phase $\theta_{1}+\theta_{2}$ is presented in Fig.~\ref{fig:XmXpN} for both non-heralded (a) and heralded (b) measurements. They correspond to states $\ket\Psi$ and $\hat a_1\hat a_2\ket\Psi$, respectively. Minimum variances $\langle (X_1 - X_2)^2 \rangle$ and $\langle (X_1 + X_2)^2 \rangle$ at local oscillator phases $\theta_{1}+\theta_{2}$ equal to $0$ and $\pi$, respectively, fall below the standard quantum limit, which indicates squeezing. We observe an increase of squeezing from $0.560 \pm 0.005$ for the undistilled to $0.83\pm0.05$ dB for the distilled  state.

For higher precision analysis, we fit the data in Fig.~\ref{fig:XmXpN}(a) by the model of quadrature variance dependence (\ref{eq:phase_dependence}) for a TMSV with the value of $\eta$ found previously. From that fit, we evaluate the initial TMSV squeezing parameter, prior to losses, as $\zeta=0.190$. If we subject this TMSV to two-mode photon annihilation operation, we obtain theoretical curves shown by solid lines in Fig.~\ref{fig:XmXpN}(b), which turn out to match well the experimental data.

Also, we fit the measured variances of the distilled state with a TMSV with an increased squeezing parameter $\zeta = 0.358$ that has undergone the same losses as the initial state [dashed lines in Fig.~\ref{fig:XmXpN}(b)]. Good agreement is obtained, indicating that the state has retained its TMSV character after distillation. 


From the quadrature measurements we reconstruct density matrices of both the initial and final states in the Fock basis up to 3 photons by means of the maximum-likelihood method \cite{Lvovsky_max_lik,DilutedMaxLik}. Absolute values of the lowest density matrix elements are presented in Fig.~\ref{fig:rho}(a,b). The off-diagonal components $|0,0\rangle \langle 1,1|$ and $|1,1\rangle \langle 0,0|$ in the final state are significantly greater than those in the  initial state, serving as evidence of higher entanglement. For a more rigorous estimation of the entanglement increase, we evaluate the log-negativity parameter $E_N$ from the reconstructed density matrixes and list it in Table~\ref{tab:log_neg}. An increase by a factor of about 50\% is present.

In an ideal setting, as seen from \eeqref{eq:psione}, we would also expect entanglement increase in the case of single-mode photon annihilation. However, in the presence of losses, the one-photon subtracted state demonstrates lower entanglement compared to initial TMSV (Table \ref{tab:log_neg}). This is because the entanglement of that state (\ref{eq:psione}) occurs due to the $|1,2\rangle$ component [Fig.~\ref{fig:rho}(c)], and the two-photon Fock state is highly sensitive to losses, more so than the single-photon state.

For fair evaluation of the distillation procedure, we should correct the parameters obtained for the unheralded state for the loss occurring in the asymmetric beam splitter. The corrected values for the squeezing and entanglement are still significantly below those for the distilled state (Table~\ref{tab:log_neg}).

The  entanglement increase factor in the procedure described in this experiment is theoretically limited by two. More entanglement can be obtained by higher-order photon annihilation, but at a cost of dramatic productivity loss. More promising techniques of CV entanglement distillation would involve nonclassical light sources or nonlinear optical interactions in both modes of TMSV. For example, a procedure involving noiseless amplification \cite{Xiang10} in both modes has no fundamental limitation on the achievable entanglement increase factor.

The experiment has been supported by NSERC and CIFAR. AL is a CIFAR Fellow. YK is an RQC Fellow. We thank Marco Barbieri, Alexey Fedorov, Joshua Nunn, Evgeniy Safonov and Ian Walmsley for helpful discussions.


\begin{table*}[ht]
\small
\caption{\label{tab:log_neg}
Parameters of the states before and after distillation}
\begin{ruledtabular}
\begin{tabular}{lccr}
\textrm{State} &
{\textrm{squeezing parameter from fit}}&{\textrm{maximum squeezing, dB}}&{log-negativity} \\
\colrule
Initial TMSV &
$0.1900\pm 0.0007$ & $0.560 \pm 0.005$ &$0.209\pm 0.002 $\\
Initial TMSV compensated \\ for beam splitter& N/A & $0.63\pm0.005$
 & $0.239 \pm 0.002$ \\
Two-mode photon subtraction & $0.358\pm 0.006$ &  $0.83\pm0.05$ &  $0.30\pm 0.01$\\
One-mode photon subtraction & N/A & N/A & $0.12\pm 0.01$ \\
\end{tabular}
\end{ruledtabular}
\end{table*}

%


\end{document}